\documentclass{ckm}                 

\usepackage{txfonts}            

\confname{Workshop on the CKM Unitarity Triangle, IPPP Durham, April
  2003}

\title{Prospects for the measurement of \Bs\ oscillations with the
  ATLAS detector at LHC}  

\author{B.~Epp\addressmark{a}, V.M.~Ghete\addressmark{a}\thanks{Speaker
    at the Workshop.}, A.~Nairz\addressmark{b}}

\address[a]{Institute for Experimental Physics, University of
  Innsbruck,  Austria} 
\address[b]{CERN, Geneva, Switzerland}

\def\EPJC#1#2#3{{Euro.~Phys.~J.~C}{\bf #1}\ (#2)\ #3}
\def\EPJdirect#1#2#3{{EPJdirect~}{\bf #1}\ (#2)\ #3}

\def\CPC#1#2#3{{Comp.~Phys.~Comm.~}{\bf #1}\ (#2)\ #3} 
\def\NIMA#1#2#3{{Nucl.~Instr.~Meth.~A}{\bf #1}\ (#2)\ #3}

\newcommand{\Bs}   {\ensuremath{B_{s}^{0}}}
\newcommand{\barBs}{\ensuremath{\bar{B}_{s}^{0}}}
\newcommand{\Bd}   {\ensuremath{B_{d}^{0}}}
\newcommand{\barBd}{\ensuremath{\bar{B}_{d}^{0}}}
\newcommand{\Ds}   {\ensuremath{D_{s}}}
\newcommand{\Dsm}  {\ensuremath{D_{s}^{-}}}
\newcommand{\Dsp}  {\ensuremath{D_{s}^{+}}}
\newcommand{\Dm}   {\ensuremath{D^{-}}}

\newcommand{\aone} {\ensuremath{a_{1}}} 
\newcommand{\aonem}{\ensuremath{a_{1}^{-}}} 
\newcommand{\aonep}{\ensuremath{a_{1}^{+}}} 
\newcommand{\pim}  {\ensuremath{\pi^{-}}} 
\newcommand{\pip}  {\ensuremath{\pi^{+}}} 
\newcommand{\Km}   {\ensuremath{K^{-}}} 
\newcommand{\Kp}   {\ensuremath{K^{+}}}  
\newcommand{\Lb}   {\ensuremath{\Lambda_{b}^{0}}} 
\newcommand{\Lcp}  {\ensuremath{\Lambda_{c}^{+}}} 

\newcommand{\ifb}  {\ensuremath{{\mathrm{fb}}^{-1}}}

\newcommand{\ips}  {\ensuremath{{\mathrm{ps}}^{-1}}}

\newcommand{\dms}  {\ensuremath{\Delta m_s}}
\newcommand{\dmd}  {\ensuremath{\Delta m_d}}

\newcommand{\dgsg} {\ensuremath{\Delta \Gamma_s /\Gamma_s}}

\newcommand{\dmsSec} {\boldmath $\Delta m_s$}

\newcommand{\pT}   {\ensuremath{p_\mathrm{T}}}

\newcommand{\etal} {\hbox{et al.}}

\newcommand{\D}    {\displaystyle}

\begin{document}

\begin{abstract}
The prospects for the measurement of $B_{s}^{0}$ oscillations with the
ATLAS detector at the Large Hadron Collider  are presented.
$B_{s}^{0}$ candidates in the $D_{s}^{-} \pi^{+}$
and $D_{s}^{-} a_{1}^{+}$ decay modes from semileptonic events 
were fully simulated and reconstructed, using a detailed detector
description. The sensitivity and the expected accuracy
for the measurement of the oscillation frequency were derived from 
unbinned maximum likelihood amplitude fits as functions of the
integrated luminosity. A detailed treatment of the systematic
uncertainties was performed. The dependence of the measurement 
sensitivity on various parameters was  also evaluated.

\end{abstract}

\maketitle


\section{Introduction}

The observed  \Bs\ and \barBs\ states are linear combinations of two mass 
eigenstates, denoted here as $H$ and $L$. Due to the non-conservation of 
flavour in charged weak-current interactions, transitions between 
\Bs\ and \barBs\ states occur with a frequency proportional to 
$\dms = m_H - m_L$.

Experimentally, the \mbox{\Bs -- \barBs} oscillations have not yet
been observed directly.  The combined lower limit from measurements
done by the ALEPH, DELPHI and OPAL experiments at LEP, by SLD at SLC,
and by CDF at the Tevatron, is  $\dms > 14.4$~\ips\ at 95\% CL, with
a sensitivity at 95\% CL of 19.3~\ips\ \cite{BoscWG}. 
In the Standard Model, it would be difficult
to accommodate values of \dms\ above $\sim 25$~\ips~\cite{CKM01}.

In this paper, the prospects of the ATLAS experiment at the Large
Hadron Collider (LHC) to measure \mbox{\Bs -- \barBs} oscillations are
presented. A detailed description of the analysis on which this
presentation is based on can be found in Ref.~\cite{BsMix_SN}. A short
discussion of subsequent changes is also included.

\section{Event selection}
\label{EvSel}

The signal channels considered in this analysis for the measurement of
\mbox{\Bs -- \barBs} oscillations are $\Bs \to \Ds \pi$ and $\Bs \to
\Ds \aone$, with $\Ds \to \phi \pi$ followed by $\phi \to \Kp\Km$,
selected in semileptonic events. 

The event samples from this simulation study were generated using
PYTHIA 5.7 \cite{PYTHIA}, passed then through a detailed GEANT3-based
simulation of the ATLAS Inner Detector; charged tracks were then
reconstructed using an algorithm based on the Kalman filter. The
production of the $b\bar{b}$-quark pairs in $pp$ collisions at a
centre-of-mass energy of $\sqrt{s} = 14$~TeV included direct
production, gluon splitting, and flavour excitation processes. The
$b$-quark was forced to decay semileptonically giving a muon with
transverse momentum $\pT > 6$~GeV and pseudo-rapidity $|\eta| < 2.5$
which is used by the level-1 trigger to select the $B$ hadronic
channels in ATLAS, while the associated $\bar{b}$ was forced to
produce the required $B$-decay channels. 

A multi-level trigger is used in ATLAS to select events. For
B-physics, the level-1 trigger is an inclusive muon trigger, as
mentioned before.  The level-2 trigger reconfirms the muon from
level-1 trigger using also the precision muon chambers, then in an
un-guided search for tracks in the Inner Detector reconstructs a
$\phi$ meson and, adding a new track, a \Ds\  meson.  The level-3
trigger (the event filter) confirms the level-2 result using a set of
loose offline cuts to select the events.

The flavour of the \Bs\ meson, i.e.\ the particle or antiparticle
state, is tagged at the production point by the muon used for the
level-1 trigger; at the decay vertex, the meson's state is given by the
charge of the reconstructed \Ds\ meson. 

Offline, the \Bs\ meson was reconstructed from its decay products,
applying kinematical cuts on reconstructed tracks, mass and vertex-fit
cuts on the intermediate particles, and cuts on properties of the \Bs\
candidates (vertex-fit quality, proper time, impact parameter, mass,
etc.).

The background was estimated considering various four- or six-body
$B$-hadron decay channels, and the combinatorial background.
The four- and six-body background events were generated, passed through the 
detailed detector simulation program, reconstructed and analyzed using the 
same programs, the same conditions and the same cuts as the signal events.
For the study of the combinatorial 
background, about 1.1~million $b\bar{b} \to \mu X$ events were analyzed 
using a fast detector simulation package.
 
The expected number of signal and background events for an integrated
luminosity of 10~\ifb, corresponding to a one-year run at
$10^{33}\;\mathrm{cm}^{-2}\mathrm{s}^{-1}$ (so-called ``low
luminosity'') are summarized in Table~\ref{tab:nev}.  To compute them,
branching ratios from Ref.~\cite{PDG1998} were taken, where known,
else from PYTHIA. Charge-conjugate channels were taken into account;
corrections for trigger efficiencies of 63\% (level-2 \Ds) and 82\%
(muons, level-1 and offline combined) were applied.

\renewcommand{\arraystretch}{1.1}
\begin{table}[htbp]
  \begin{center}
    \begin{tabular}{llr}
      \hline
                                                  &
      \multicolumn{1}{l}{Process}                 &
      \multicolumn{1}{r}{Events}                  \\ \hline
      Signal                                      &
      $\Bs \to \Dsm \pip$                         & 
      2370                                        \\ 
      channels                                    & 
      $\Bs \to \Dsm \aonep$                       & 
      870                                         \\ \hline 
                                                  &
      $\Bd \to \Dsp \pim$                         &
      $<$ 400                                     \\
      Exclusive                                   &
      $\Bd \to \Dsp \aonem$                       &
      $<$ 340                                     \\
      background                                    &
      $\Bd \to \Dm  \pip$                         &
      3                                           \\
      channels                                    &
      $\Bd \to \Dm  \aonep$                       &
      1                                           \\
                                                  &
      $\Lb \to \Lcp (p \Km \pip) \pim$            &
      2                                           \\
                                                  &
      $\Lb \to \Lcp (p \Km\, 3\pi) \pim$          &
      0                                           \\ \hline
      Combin.                                     &
      4 charged tracks                            &
      1920                                        \\
      background                                    &
      6 charged tracks                            &
      1830                                        \\ \hline
    \end{tabular}
    \caption{{\small Signal and background samples analyzed for the
             study of \Bs -- \barBs\  oscillations, and numbers
             of events expected for 10 \ifb.}}
    \label{tab:nev}
  \end{center}
\vspace*{-5.5ex}
\end{table}
\renewcommand{\arraystretch}{1.0}

The reconstructed \Bs\ invariant-mass distribution in the decay
channel $\Bs \to \Dsm \pip$ is shown in Figure~\ref{fig:mass_Dspi}
for an integrated luminosity of 10~\ifb.

\vspace*{-5.5ex}
\begin{figure}[htbp]
  \centering \hbox to\hsize{\hss
    \includegraphics[width=\hsize]{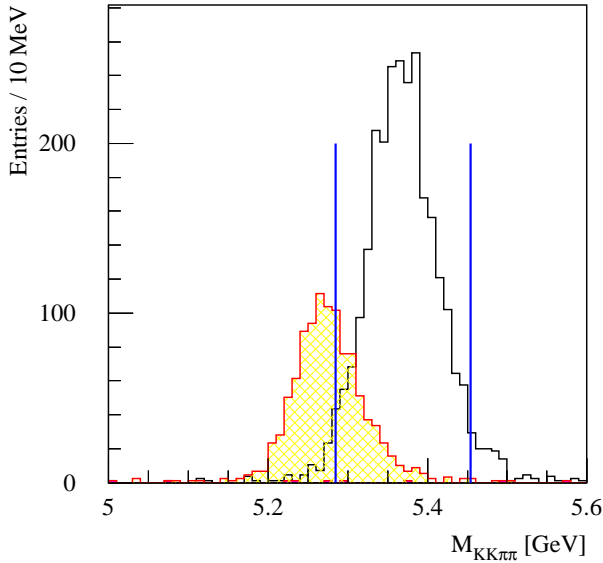} 
    \hss}
    \caption{Reconstructed \Bs\ invariant-mass distribution for $\Bs \to \Dsm
      \pip$ decays. Open histogram: signal, hatched: background from
      $\barBd \to \Dsm \pip$ decays, dark: fake reconstructed decays
      from the signal sample.  Combinatorial background not shown
      here.}
    \label{fig:mass_Dspi}
\end{figure}

\section{Extraction of significance limits and accuracy for
      the measurement of \dmsSec}
\label{dmsLimAcc}

\subsection{Proper-time reconstruction and resolution }
\label{pTime}

The proper time of the reconstructed \Bs\  candidates was computed from 
the reconstructed transverse decay length, $d_{xy}$, and from the \Bs\ 
transverse momentum, \pT:
\begin{displaymath}
t = \frac {d_{xy} M_{\Bs}} 
          {c \pT}
 \equiv  d_{xy} g
\end{displaymath}
where $g = M_{\Bs}/(c \pT)$ and $M_{\Bs}$ is the \Bs\ mass.  Its
resolution function $\mathrm{Res}(t \, | \, t_0)$ was
pa\-ram\-e\-ter\-ized with the sum of two Gaussian functions, see
Eq.~(\ref{eq:2G_resfunc}),  with
parameters given in Table~\ref{tab:ptres2G_par} for the signal
channels. Here $t_{0}$ denotes the 
true (generated) proper time; $f_\alpha$ is the fraction and 
$\sigma_{\alpha}$ the width of the Gaussian function $\alpha$.
Similar parameterizations were obtained for the background
channels $\Bd\to\Dsp\pim$ and $\Bd\to\Dsp\aonem$.
\begin{eqnarray}
    \mathrm{Res}(t \, | \, t_0) \!\!\!
    & = & \!\!
    f_{1} \frac{1}{\sigma_{1}\sqrt{2\pi}}\;\,
    \exp\left( -\frac{(t-t_0)^{2}}{2\,\sigma_{1}^{\; 2}}\right)
    + \nonumber \\     
    \!\!\! &   & \!\!
    f_{2} \frac{1}{\sigma_{2}\sqrt{2\pi}}\;\,
    \exp\left( -\frac{(t-t_0)^{2}}{2\,\sigma_{2}^{\; 2}}\right)
  \label{eq:2G_resfunc}
\end{eqnarray}

\setlength{\tabcolsep}{4.5pt}
\begin{table}[htbp]
\begin{center} 
  \begin{tabular}{ccccc} \hline
            &  \multicolumn{2}{c}{$\Bs \to \Dsm\,\pi^+$}   
            &  \multicolumn{2}{c}{$\Bs \to \Dsm\,a_1^+$}  \\ \cline{2-5} 
  $\alpha$  &  \multicolumn{1}{c}{$f_{\alpha}$ (\%)}
            &  \multicolumn{1}{c}{$\sigma_{\alpha}$(fs)}
            &  \multicolumn{1}{c}{$f_{\alpha}$ (\%)}
            &  \multicolumn{1}{c}{$\sigma_{\alpha}$(fs)}  \\ \hline      
1 & $59.6 \pm  6.6$  & $\phantom{1}51.5 \pm  4.0$ & 
    $62.5 \pm 14.1$  & $\phantom{1}51.6 \pm  \phantom{1}6.4$
                                                                  \\ 
2 & $40.4 \pm  6.6$  & $          107.3 \pm  8.5$ & 
    $37.5 \pm 14.1$  & $\phantom{1}92.8 \pm 12.7$ 
                                                                  \\ \hline
  \end{tabular}
  \caption{Proper-time resolution function $\mathrm{Res}(t \, | \, t_0)$ 
           parameterization with the sum of two Gaussian functions.} 
  \label{tab:ptres2G_par}
\end{center}
\end{table}
\setlength{\tabcolsep}{6pt}

\subsection{Likelihood function}
\label{LL}

The probability density to observe an initial $B_{j}^{0}$  meson ($j=d, \;s$)
decaying at time $t_0$ after its creation as a $\bar{B_{j}^{0}}$
meson is  given by:
\begin{eqnarray}
    p_j(t_0, \; \mu_0) \!\!\!
    & = & \!\! 
     \D \frac{\Gamma_{j}^{2} - \left( \frac{\Delta\Gamma_{j}}{2}\right)^{2}}
    {2\,\Gamma_{j}} \: \mathrm{e}^{- \Gamma_{j} t_0} \; \times  \nonumber \\
    \!\!\! &  & \!\!
    \left( \cosh \frac{\Delta\Gamma_{j} t_0}{2} + \mu_0 \cos\Delta m_{j} t_0 
    \right) 
  \label{eq:oscPr}
\end{eqnarray}
where $\Delta\Gamma_{j}=\Gamma_\mathrm{H}^{j}-\Gamma_\mathrm{L}^{j}$, 
$\Gamma_j = (\Gamma_\mathrm{H}^{j} + \Gamma_\mathrm{L}^{j})/2$ and 
$\mu_0 = -1$. 
For the unmixed case (an initial  $B_{j}^{0}$ meson decays as a $B_{j}^{0}$ 
meson at time $t_0$), the probability density 
is given by the above expression with $\mu_0 = +1$. 

The above probability is modified by experimental effects: finite
proper-time resolution, wrong tags at production or decay, and
background. Convolving  $p_j(t_0, \; \mu_0)$ with the proper 
time resolution $\mathrm{Res}_j(t \, | \, t_0)$, one obtains the
probability as a function of $\mu_0$ and the reconstructed proper 
time~$t$:
\mbox{$
  q_j(t, \mu_0)= N 
  \int_{t_\mathrm{min}}^\infty  
       p_j(t_0,\mu_0)\; \mathrm{Res}_j(t\; |\; t_0) \; \mathrm{d}t_0 
$}, 
with $N$ a normalization factor and $t_\mathrm{min} =0.4$~ps the
cut on the \Bs\ proper decay time.
Assuming a fraction $\omega_j$ of wrong tags at production or decay, the 
probability becomes
\mbox{$
  q_j^\prime(t, \mu)= (1-\omega_j) q_j(t,\mu) + \omega_j q_j(t,-\mu) 
\label{eq:qjt}
$}. 
Including the background, composed of oscillating \Bd\ mesons
and of combinatorial background, with fractions $f_j^k$ 
($j=$ $s$, $d$, and com\-bi\-na\-to\-ri\-al background $cb$),
one obtains:  
\mbox{$
  \mathrm{pdf}_k(t,\mu) = 
  \sum_{j=s,d,cb} f_j^k \left  [ (1-\omega_j) q_j(t,\mu) + \omega_j q_j(t,-\mu) 
                        \right ] 
$}
where the index $k=1$ denotes the $\Bs \to \Dsm \pip$ 
channel and $k=2$ the $\Bs \to \Dsm \aonep$ channel.
The likelihood of the total sample is written as
\begin{equation}
{\cal L}(\Delta m_s, \Delta \Gamma_s) = 
    \prod_{k=1}^{N_{\mathrm{ch}}^{ }} \;
    \prod_{i=1}^{N_{\mathrm{ev}}^k} \mathrm{pdf}_k(t_i,\mu_i)
\label{fun:LL}
\end{equation}
where $N_{\mathrm{ev}}^k$ is the total number of events of  type $k$, and 
\mbox{$N_{\mathrm{ch}} = 2$}. 

\subsection{Significance limits  for the measurement of \dmsSec}
\label{dmsLim}

The ATLAS sensitivity for the  \dms\ measurement was determined using a
simplified Monte-Carlo model to produce event samples, combined with
the amplitude-fit method~\cite{AFit} to extract the limits.  In the
amplitude-fit method a new parameter, the \Bs\ oscillation amplitude
${\cal A}$, is introduced in the likelihood function by replacing the
term `$\mu_0 \cos\Delta m_{s} t_0$' with `$\mu_0 {\cal A} \cos\Delta
m_{s} t_0$' in the \Bs\ probability density function.  For each value
of \dms, the new likelihood function is minimized with respect to
${\cal A}$, keeping all other parameters fixed, and a value ${\cal A}
\pm \sigma_{\cal A}^{\mathrm{stat}} $ is obtained. The statistical
significance $S$ of an oscillation signal can be expressed as $S
\approx 1/\sigma_{\cal A}$.  One defines a $5\sigma$ significance limit
as the value of \dms\ for which $1/\sigma_{\cal A} =5$, and a
sensitivity at 95\% confidence limit as the value of \dms\ for which
$1/\sigma_{\cal A} = 1.645$.

For \dms\ values smaller than the $5\sigma$ significance limit, the expected 
accuracy is estimated using the  log-likelihood method, with the likelihood 
function given by Eq.~(\ref{fun:LL}).

\subsection{Systematic uncertainties}
\label{dmsSys}

An attempt to estimate the systematic uncertainties was done. The
following contributions to the systematic uncertainties were
considered: a relative error of 5\% on the wrong-tag fraction for both
\Bs\ and \Bd; $\pm 1\sigma$ variation of Gaussian-function widths from
$\mathrm{Res}(t \, | \, t_0)$ parameterization; $f_{\Bs} = BR (\bar{b}
\to \Bs)$, \Bs\ lifetime, \dmd\ varied separately by PDG uncertainty;
5\% uncertainty for decay time $\tau_{\mathrm{cb}}$ of
combinatorial background, keeping the shape exponential. An additional
set of `projected systematic uncertainties' was defined,  reducing 
the  $f_{\Bs}$ error and the uncertainties on the widths from the
proper time parameterization to values expected at the time of ATLAS
data taking. 

\section{Results and conclusions}
\label{results}

Table~\ref{tab:syst_dms} shows the dependence of the amplitude and its
statistical and systematic uncertainties on \dms\ for an
integrated luminosity of 10~\ifb, for both actual and projected
systematic uncertainties. In the generated event samples, the value of
$\dms$ was set to $\dms^{\mathrm{gen}} = \infty$, therefore one
expects $\sigma_{\cal A}$ compatible with zero.  The dominant
contributions to the systematic uncertainty come from the uncertainty
on the $f_{\Bs}$ fraction and from the parameterization of the proper time
resolution.

\begin{table}[htbp]
  \begin{center}
    \begin{tabular}{lrrrr} 
       \hline
       \multicolumn{1}{c}{\dms}       &  
        0 \ips\  &  
       10 \ips\  & 
       20 \ips\  & 
       30 \ips\  \\ \hline
       \multicolumn{1}{c}{${\cal A}$} &  
       $ 0.045$  &
       $ 0.189$  &          
       $ 0.042$  &          
       $-0.291$  \\ \hline
       \multicolumn{1}{c}{$\sigma_{\cal A}^{\mathrm{stat}}$} &  
       $\pm 0.048$  &         
       $\pm 0.090$  &
       $\pm 0.167$  &
       $\pm 0.357$  \\ \hline
       \multicolumn{1}{c}{$\sigma_{\cal A}^{\mathrm{syst}}$} &  
       $^{+0.097\rule[1ex]{0mm}{0.5ex}}_{-0.084\rule[-0.5ex]{0mm}{0.5ex}}$ &
       $^{+0.130}_{-0.096}$ &
       $^{+0.180}_{-0.142}$ &
       $^{+0.298}_{-0.226}$ \\ \hline
       \multicolumn{5}{c}{with `projected systematic uncertainties'} \\ \hline  
       \multicolumn{1}{c}{$\sigma_{\cal A}^{\mathrm{syst}}$}  &  
       $^{+ 0.049\rule[1ex]{0mm}{0.5ex}}_{-0.049\rule[-0.5ex]{0mm}{0.5ex}}$ &
       $^{+ 0.060}_{-0.048}$ &
       $^{+ 0.085}_{-0.066}$ &
       $^{+ 0.137}_{-0.117}$ \\ \hline
    \end{tabular}
    \vspace*{0.5ex} 
    \caption{\small The oscillation amplitude ${\cal A}$ and its statistical and 
      systematic uncertainties as a function of \dms\ for an integrated 
      luminosity of 10~\ifb.}    
    \label{tab:syst_dms}
  \end{center}
  \vspace*{-6ex} 
\end{table}

\begin{figure}[htbp]
  \centering \hbox to\hsize{\hss
    \includegraphics[width=\hsize]{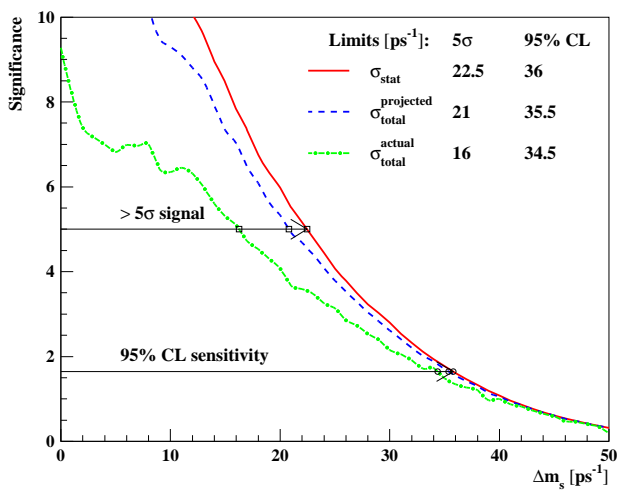} 
    \hss}
  \caption{The significance of the \Bs\ oscillation signal as a
    function of \dms\ for an integrated luminosity of 10~\ifb.}  
  \label{fig:sign_dms}  
\end{figure}

The significance  of the \Bs\ oscillation signal as a function of \dms\ for 
an integrated luminosity of 10~\ifb is shown in
Fig.~\ref{fig:sign_dms}. The $5\sigma$ significance limit is
22.5~\ips\ and the 95\% CL sensitivity 
is 36.0~\ips, when computed with the statistical uncertainty only.
Computed with the total uncertainty, the $5\sigma$ significance limit is 
16.0~\ips\ and the 95\% CL sensitivity is 34.5~\ips\ for the actual
systematic uncertainties, and 21~\ips\ and 35.5~\ips\ for the
projected systematic uncertainties. The limits for other values of the
integrated luminosity are given in  Table~\ref{tab:dms_lum}, computed with the
statistical uncertainties only. 

\begin{table}[htbp]
  \begin{center}
    \begin{tabular}{ccc} 
\hline
Luminosity   & $5 \sigma$ limit         &   $95\%$ CL sensitivity  \\ 
  (\ifb)     &   (\ips)                 &          (\ips)          \\ 
\hline
\phantom{1}5 &     17.5                 &      32.0                \\          
          10 &     22.5                 &      36.0                \\       

          20 &     27.0                 &      39.0                \\  
          30 &     29.5                 &      41.0                \\       
\hline
    \end{tabular}
    \caption{The dependence of significance limits for the \dms\
          measurement on the integrated luminosity, computed with 
          statistical uncertainties only.}     
    \label{tab:dms_lum}
  \end{center}
\end{table}

The dependence of the significance limits on \dgsg\  was also
estimated. For values of \dgsg\ up to 30\%, no sizeable effect was
observed. The shape and the fraction of the combinatorial background
were also varied within reasonable values; only a weak dependence of
the limits was observed.

For \dms\ values smaller than the $5\sigma$ significance limit, the
accuracy of the \dms\ measurement was determined for different 
values of the integrated luminosity. The results are given in
Table~\ref{tab:dmsAcc}. If measured, the precision on 
\dms\ will be dominated by the statistical errors.

\setlength{\tabcolsep}{4.9pt}
\begin{table}[htbp]
  \begin{center}
    \begin{tabular}{cccc} 
\hline
Luminosity                                      &
$\dms^{\mathrm{gen}}$                           & 
$\dms^{\mathrm{rec}} \pm \sigma^{\Delta m_s}_{\mathrm{stat}} 
                     \pm \sigma^{\Delta m_s}_{\mathrm{syst}}$  &   
Obs.                                                        \\ 
  (\ifb)        & (\ips) &      (\ips)                  &   \\ 
\hline
\phantom{1} 5   & $17.5$ & $17.689 \pm 0.083 \pm 0.002$   & $5 \sigma$ lim. \\
\hline 
                & $15.0$ & $15.021 \pm 0.049 \pm 0.002$   & \\ 
\raisebox{1.5ex}[0cm][0cm]{10}   
                & $22.5$ & $22.396 \pm 0.072 \pm 0.005$   & $5 \sigma$ lim. \\
\hline
                & $15.0$ & $14.949 \pm 0.033 \pm 0.002$   & \\ 
           20   & $20.0$ & $20.041 \pm 0.068 \pm 0.005$   & \\ 
                & $27.0$ & $26.948 \pm 0.070 \pm 0.003$   & $5 \sigma$ lim. \\
\hline
                & $15.0$ & $14.942 \pm 0.028 \pm 0.004$   & \\ 
           30   & $20.0$ & $20.010 \pm 0.043 \pm 0.002$   & \\ 
                & $29.5$ & $29.708 \pm 0.083 \pm 0.007$   & $5 \sigma$ lim. \\
\hline
    \end{tabular}
    \caption{The accuracy of \dms\ measurement as a function of the 
             integrated luminosity. $\sigma^{\Delta m_s}_{\mathrm{stat}}$
             represents the statistical uncertainty, 
             $\sigma^{\Delta m_s}_{\mathrm{stat}}$ the systematic
             uncertainty. The $5 \sigma$ limits were computed with
             statistical errors only.}
    \label{tab:dmsAcc}
  \end{center}
  \vspace*{-6ex} 
\end{table}
\setlength{\tabcolsep}{6pt}

\section{Recent developments}
\label{devel}  

This section summarizes recent changes of the assumptions or
conditions used to get the results presented above. 

The most important changes in the detector geometry are the increase of
the beam-pipe diameter from 41.5~mm to 50.5~mm and the increase of the
pixel length in B-layer, the closest layer to the beam-pipe, from
300~${\mu}$m to 400~${\mu}$m.
 
Due to financial constraints, the  B-physics trigger resources have to
be minimized. Previously, dedicated resources were supposed to be
available for the B-physics trigger, in addition to resources for
high-\pT\ physics trigger (`discovery physics trigger'). It may not be
possible  to provide any significant additional resources. 
Moreover, there are financial uncertainties which could lead to the
deferral of some detector items, therefore  it is possible to  
have a reduced detector at start-up. Items included in the deferral
scenario are the second pixel layer from the Inner Detector, resulting
in a two-layer pixel detector, the $\eta > 2$ region of the Transition 
Radiation Detector from the Inner Detector, and a significant part of
the processors for level-2 and event filter, reducing the computing
resources for the high-level trigger and limiting the level-1 rate. 

The luminosity target for LHC start-up doubled to $2 \times
10^{33}$~cm$^{-2}$s$^{-1}$, therefore it will be necessary to
re-evaluate trigger thresholds and to remove some triggers requiring
too much resources. The trigger for \Bs\ oscillation channels requires a
significant rate, therefore it is very likely that the muon trigger 
threshold will be raised to $\sim 8$~GeV. 

Recent work concentrates on trigger-related issues; improvements of
the offline analysis, although possible, have to be postponed.
To reduce the resource requirements, one of the possibilities would be  
to change the trigger for $B$ hadronic channels from  
$\mu (\pT > 6\;\mathrm{GeV})$ at level-1 plus Inner Detector full scan
at Level-2 to $\mu (\pT > 6\;\mathrm{GeV})$ at level-1, plus a low $E_T$
level-1 calorimeter Region-of-Interest (RoI),  used then to guide
reconstruction at level-2. In addition, one can use the level-2 RoI to
limit the region for reconstruction at the event filter.
A flexible trigger and analysis strategy is nowadays evaluated,
which should be able to cope with detector changes, luminosity
scenarios, financial problems. Another direction is to recover
performance using optimized reconstruction algorithms, flexible
trigger and analysis thresholds. A re-evaluation of the detector
performance with the latest geometry is also under way. Preliminary
results are encouraging.

\end{document}